# Physically Informed Data-Driven Modeling of Low-Pressure Carburizing (LPC): A Self-Calibrating Boundary Condition for Carbon Flux

**Oleksii Sagalovych**[1] - ORCID https://orcid.org/0000-0003-2136-2740
**Tymofii Sagalovych**[2] - ORCID https://orcid.org/0009-0006-7944-8079
**Stanislav Dudnik**[1] - ORCID https://orcid.org/0000-0003-2074-4739
**Vladislav Sagalovych**[1] - ORCID https://orcid.org/0000-0002-8060-3201

***Abstract.*** *This paper formulates an extended framework for a hybrid physically informed data-driven model of low-pressure carburizing (LPC), in which diffusive carbon transport in austenite is described by a physical equation, whereas the effective carbon flux across the gas-metal interface is treated as an unknown boundary forcing to be identified from experimental and process data. The basis for this formulation is that, in LPC, the most uncertain part of the model is not bulk diffusion but the transient surface kinetics of carbon uptake by the metal. In contrast to direct empirical relationships of the "process regime-result" type, the proposed scheme uses machine learning as a closure operator for the boundary condition in the diffusion problem. This operator is calibrated using carbon concentration profiles, integral mass gain, surface concentration, effective case depth, and additional process measurements. Theoretically, the model is positioned as an operator-learning problem in a system constrained by a partial differential equation, physical flux constraints, inverse-problem regularization, and uncertainty quantification. The relationships between the proposed approach and physics-informed neural networks, operator learning, dynamic-system identification, boundary-operator learning, and boundary-control problems are considered. The mathematical formulation, calibration functional, identifiability conditions, preliminary tuning algorithm, and online adaptation scheme are formulated. The proposed approach is considered as a computational basis for developing interpretable digital support for LPC processes.*

## 1. Introduction

Surface engineering plays a key role in improving the wear resistance, contact fatigue strength, and durability of structural steels without the need to change the composition and properties of the entire component volume. Among industrial methods of thermochemical hardening, processes that enrich the surface layer with carbon and nitrogen, including carburizing, carbonitriding, and nitriding, are of particular importance. Their technological objective is to form a surface zone with increased hardness while retaining a tougher core, which is especially important for gears, shafts, bearing components, and other elements operating under contact and cyclic loading [1]–[5].

[1] *JSC "FED", Kharkiv, Ukraine*
[2] *University of Applied Sciences and Arts, Lucerne, Switzerland, https://ror.org/04nd0xd48*
✉ *A. V. Sagalovych*
*iht@kharkov.ua*

Low-pressure carburizing (LPC) is a modern development of vacuum carburizing. Unlike atmospheric gas carburizing, where process control is usually associated with the carbon potential of the atmosphere, LPC is carried out in a rarefied environment with pulsed supply of a hydrocarbon gas, most often acetylene or another active carbon-containing gas. The process usually includes alternating stages of active enrichment and diffusion holds, which makes it possible to form a specified carbon profile while reducing the risk of intergranular oxidation and improving surface cleanliness [2]–[6].

The main difficulty in modeling LPC is that, unlike bulk diffusion of carbon in austenite, which has a sufficiently developed physical basis, the surface kinetics of carbon uptake remains weakly observable. The actual carbon flux across the gas-metal interface depends on temperature, pressure, gas-phase composition, gas flow rate, surface condition, steel grade, load area and packing density, and the history of previous active-enrichment stages and diffusion holds [2], [5]–[9]. Therefore, even a correct diffusion model may produce an erroneous result if the flux boundary condition is specified incorrectly.

The aim of the present work is to formulate a hybrid model in which the physical core is retained as a diffusion equation, whereas the most uncertain part of the process, namely the carbon flux through the surface, is described by a self-calibrating boundary operator. In this formulation, machine learning does not replace the physical model but is used to close the boundary condition in the inverse problem of identifying a hidden surface forcing. This makes it possible to preserve the physical interpretability of the calculation and, at the same time, to use data to adapt the model to a particular steel, load, and process regime.

## 2. The LPC Process as a Modeling Object

LPC can be considered as a transient thermochemical process in which the gas phase, the metal surface, and bulk diffusion are linked by a single transport chain. During the active-enrichment stage, the hydrocarbon gas enters the working chamber, decomposes on the heated steel surface, and generates a carbon flux into the metal. During the diffusion-hold stage, the supply of active gas is stopped or substantially reduced, and the carbon introduced earlier is redistributed into the depth by diffusion in austenite [3]–[6].

The final technological result is determined not only by the effective depth of the carburized layer. The surface carbon concentration, the shape of the concentration profile, the uniformity of enrichment, the risk of carbide formation and carbonaceous deposits, and the subsequent cooling regime and formation of a martensitic structure after quenching are also important [6]–[8]. Consequently, the computational model should predict not a single integral indicator but the concentration profile and the related indicators of technological risk.

In the classical diffusion formulation, the carbon concentration profile is determined by the transport equation in austenite. For LPC, however, the question of what carbon flux actually enters through the surface during each active-enrichment stage becomes central. The instantaneous carbon flux is not directly measured in most industrial processes; only indirect data are available: the final concentration profile, surface carbon, mass gain, case depth, microstructural signs of supersaturation, and furnace-cycle parameters [2], [5], [6].

## 3. Problems to Be Solved

The first problem is the hidden nature of the boundary flux. The researcher usually has only the process results, whereas the function $J_C(t)$, which describes carbon uptake through the surface, remains unknown. Therefore, its reconstruction is an inverse problem with respect to diffusive transport.

The second problem is associated with nonstationarity. The carbon flux during a single active-enrichment stage does not have to be constant. At the beginning of the stage, the surface may accept carbon intensively, but as the surface concentration increases and the state of the gas-metal interface changes, the efficiency of further transfer decreases. Averaging the flux over the entire stage loses this information and may lead to regimes with local supersaturation [2], [6].

The third problem is the ambiguity of identification. The same final profile may be partially explained by a change in the diffusion coefficient, a temperature error, an incorrect estimate of the load area, a change in the boundary flux, or measurement noise. Therefore, the reconstruction of $J_C(t)$ requires an independently specified diffusion part, physical constraints, and regularization.

The fourth problem is transferability. A model calibrated for one steel, one furnace, or one cycle structure may lose accuracy when the steel composition, load density, pressure, gas flow rate, or stage duration is changed. Consequently, the boundary operator should be trained not as a table of regimes but as a functional dependence that accounts for process parameters and the surface state.

## 4. Problem Statement

The surface layer of a steel component is considered in the domain $\Omega = [0, L]$, where $x = 0$ corresponds to the surface and $x = L$ is chosen greater than the expected depth of the carburized layer. The unknown quantity is the carbon concentration $C(x, t)$. The physical core of the model is written as a nonlinear diffusion equation:

$$\frac{\partial C}{\partial t} = \frac{\partial}{\partial x}\left[ D(C,T,\boldsymbol{X}) \frac{\partial C}{\partial x} \right],$$

where $D(C,T,\boldsymbol{X})$ is the diffusion coefficient of carbon in austenite, $T$ is the process temperature, and $\boldsymbol{X}$ is the vector of the chemical composition of the steel [10]–[13]. The initial concentration is specified as $C(x,0) = C_0$, and $C(L,t) = C_0$ may be imposed at the remote boundary.

The unknown part of the model is concentrated at the component surface. The boundary condition is written in flux form:

$$-\rho D(C,T,\boldsymbol{X}) \left.\frac{\partial C}{\partial x}\right|_{x=0} = J_C(t),$$

where $\rho$ is the density of steel and $J_C(t)$ is the mass flux of carbon through the surface. In the conventional formulation, $J_C(t)$ must be known in advance or specified empirically. In the proposed formulation, it is determined by a trainable boundary operator:

$$J_C(t) = u(t) \cdot K_\theta(\boldsymbol{z}_t)\left[ C_{\lim}(T,\boldsymbol{X}) - C_s(t) \right]_+,$$

where $u(t)$ describes the presence of an active-enrichment stage, $C_s(t) = C(0,t)$ is the surface concentration, $\boldsymbol{z}_t$ is the vector of process parameters and states, $K_\theta$ is the trainable effective carbon-transfer coefficient, and $\theta$ denotes the model parameters. The objective is to identify $K_\theta$ from data so that the solution of the diffusion problem is consistent with experimental observations and physical constraints.

## 5. Literature Context and Theoretical Positioning

### 5.1. LPC and Carbon-Flux Models

Existing LPC models can be divided into diffusion models, models of active-enrichment/diffusion-hold cycles, and carbon-flux models [2]–[6]. Diffusion models describe bulk carbon transport well but require the surface flux to be specified correctly [10]–[13]. Boost-diffusion models are closer to the real technology because they take into account the pulsed nature of the process [3], [4]. Studies of carbon-flux parameters show that $J_C(t)$ is the critical parameter governing the accuracy of low-pressure vacuum carburizing control [2], [5], [6].

This group of studies implies that the carbon flux is not only a parameter of the gas atmosphere but also the result of interactions among the gas phase, the surface, the current carbon concentration, and the diffusion capacity of the metal. Therefore, the boundary condition should not be treated as an external constant that is completely independent of the state $C_s(t)$.

### 5.2. Machine Learning in Heat Treatment and Thermochemical Treatment

Data-analysis methods in heat treatment are used for selecting regimes, approximating relationships between process parameters and properties, classifying microstructural states, and optimizing technological cycles [1], [14]–[16]. Their advantage lies in the ability to describe complex nonlinear dependencies when analytical knowledge of the mechanism is incomplete. However, a direct "process parameters-result" model poorly separates physical causes: the contribution of flux, the contribution of diffusion, the contribution of cooling, and the contribution of measurement noise are mixed within a single empirical mapping.

For LPC, the more rigorous approach is not to replace the entire model with a neural network but to localize the trainable part in the region where the physical description is genuinely incomplete. That region is the boundary kinetics. In the proposed scheme, machine learning plays the role not of a general predictor but of a structural closure for the surface flux.

### 5.3. Physics-Informed Learning and Inverse Problems

Physics-informed neural networks have formed a general approach in which a neural approximation is trained not only on data but also on the residual of a differential equation and on initial and boundary conditions [14], [15]. For the present work, what is important is not so much the possibility of replacing the numerical solver with a neural network as the idea of jointly using data and transport laws. In problems with incomplete observations, physical constraints reduce the space of admissible solutions and act as a regularizer.

The proposed model differs from the classical PINN scheme. Here, the neural network is not required to approximate the entire function $C(x, t)$. The solution of the diffusion equation may be computed by a conventional numerical method, while only the boundary operator $K_\theta$ remains trainable. Therefore, the model is closer to learning under constraints imposed by a partial differential equation: the parameters θ are optimized through the solution of the forward problem, and the error functional is defined on observable outputs. This formulation corresponds to the inverse problem of reconstructing an unknown boundary forcing.

### 5.4. Operator Learning and the Functional Nature of the Boundary Condition

Operator learning extends classical machine learning from finite-dimensional vectors to mappings between function spaces. DeepONet is based on the universal approximation of nonlinear operators [17], and neural operators, including the Fourier Neural Operator, develop the idea of learning mappings between functions and solutions of parametric partial differential equations [18], [19]. For the problem considered here, this is fundamentally important: the carbon flux is not a single number but a function of time that depends on the history of the regime and the surface state.

Formally, the trainable boundary closure can be considered as an operator:

$$\mathfrak{B}_\theta : \left(h_t, T, p, g, q, \boldsymbol{X}, A, S, C_s\right) \rightarrow J_C\left(t\right).$$

Unlike a full neural operator that could directly map a regime to the profile $C(x, t_f)$, it is proposed to train not the entire solution operator but only its boundary component. This reduces the dimensionality of the learning problem, preserves physical interpretability, and makes it possible to control the mass balance.

It should be emphasized that DeepONet, the Fourier Neural Operator, Bayesian physics-informed networks, and related architectures are not considered in this work as already proven optimal tools specifically for LPC. They define possible classes of approximation for the trainable boundary operator. The final choice of architecture should be determined by the size of the experimental database, the noise level, the required interpretability, the need for uncertainty quantification, and the computational limitations of a specific technological application.

### 5.5. Uncertainty Quantification and Bayesian Interpretation

In problems with noisy and incomplete data, a point prediction of the boundary flux is insufficient. Bayesian variants of physics-informed neural networks and probabilistic-learning methods show that uncertainty can be introduced as part of the model itself [20], [21]. For LPC, this is particularly important: the experimental database may be sparse, and data on the surface, the load, and the gas environment may be incomplete.

In the proposed scheme, uncertainty should be associated not only with the final profile $C(x, t_f)$ but also with the reconstructed $J_C(t)$. This may be implemented using various probabilistic approaches that enable the estimation of posterior distributions for both predicted quantities and model parameters. When the model is applied outside the domain in which it was calibrated, an increase in uncertainty should be interpreted as an indicator of reduced predictive reliability.

## 6. Scientific Novelty and Boundaries of Originality

The novelty of the proposed approach does not lie in the use of artificial intelligence as such or in the use of a diffusion equation for carburizing. These elements are known from the literature. The novelty lies in the structural localization of the trainable part of the model: machine learning is applied to identify a physically constrained boundary operator of carbon flux, rather than to directly predict the final result.

This approach can be defined as a hybrid closure of an inverse boundary problem. The diffusion equation remains the transport operator, while the trainable boundary law determines what flux must be supplied to the surface so that the calculated profile is consistent with the experiment. The degree of novelty is moderate at the level of formulation and may become substantial if the transferability of the operator across different regimes, steels, and loads is experimentally confirmed.

The approach should not be presented as a completed industrial model without validation. Its testable contribution is the hypothesis that the most rational place for machine learning in an LPC model is at the gas-metal boundary rather than within the diffusion volume.

To prevent the known and the proposed contributions from being conflated, the contribution can be presented as follows (Table 1).

## 7. Mathematical Formulation of the Hybrid Model

### 7.1. Forward Problem and Solution Operator

Let $V$ be a functional space of admissible concentration profiles, for example a subspace of $H^1(0,L)$, and let $U$ be the space of admissible boundary fluxes. For a prescribed boundary flux $J_C \in U$, the diffusion equation defines the solution operator:

$$\mathcal{S}_D : J_C(t) \to C(x, t_f).$$

The observable experimental quantities are defined by an observation operator $\mathcal{O}$, which may include discrete profile values, surface concentration, effective case depth, and integral mass gain:

$$y = \mathcal{O}\left(\mathcal{S}_D(J_C)\right) + \varepsilon,$$

where $\varepsilon$ denotes measurement noise. The identification of the boundary flux is an inverse problem: it is necessary to find $J_C$ or a parameterized operator $\mathfrak{B}_\theta$ that makes the model consistent with the data.

### 7.2. Adaptive Boundary Operator

The boundary flux is given in the form:

$$J_C(t) = u(t) K_\theta(\boldsymbol{z}_t)\left[C_{lim}(T, \boldsymbol{X}) - C_s(t)\right]_+ .$$

Here, $K_\theta(\boldsymbol{z}_t)$ must be nonnegative. For this purpose, it is convenient to use the parameterization

$$K_\theta(\boldsymbol{z}_t) = \log\left(1 + \exp\left(\mathfrak{B}_\theta(\boldsymbol{z}_t)\right)\right).$$

This form ensures a physically admissible direction of flux during the active-enrichment stage and automatically weakens carbon uptake as the surface approaches the limiting concentration. If necessary, $K_\theta$ can be implemented as regularized regression, a Gaussian process, an ensemble of trees, a neural network, a recurrent history model, or a compact operator block.

### 7.3. Relationship with Boundary Control

From a mathematical point of view, $J_C(t)$ can be interpreted as boundary control in a parabolic equation. However, the objective of this work is not optimal control of the flux as a free function. The flux must be consistent with physically realizable technological parameters: gas, pressure, flow rate, temperature, load, and surface state. Therefore, the trainable operator is not an arbitrary controller but an identified technological map between the regime and the boundary forcing.

**Table. 1.** The role of machine learning in the LPC model

| Element | Status in the literature | Role in the proposed work |
|---|---|---|
| Carbon diffusion in austenite | Known physical basis for carburizing modeling | Retained as the physical core of the model |
| Carbon flux in LPC | Known as a key parameter of control and modeling | Treated as a hidden boundary forcing |
| Machine learning in heat treatment | Used for property prediction and regime optimization | Localized in the boundary operator rather than in the whole model |
| Physics-informed learning | Known methodology for forward and inverse problems | Used as a theoretical framework for constrained learning |
| Operator learning | Known class of methods for approximating functional mappings | Considered as a possible class of models for $K_\theta$ |

## 8. Identifiability, Regularization, and Stability

Identifiability of the boundary operator is a central condition for the applicability of the model. If $D(C, T, \boldsymbol{X})$, $K_\theta$, and $C_{lim}$ are allowed to vary freely at the same time, the problem becomes ill-conditioned: different combinations of these quantities may produce similar final profiles. Therefore, the diffusion part should be specified independently or calibrated on a separate dataset.

To improve identifiability, it is necessary to use not only the final case depth but also data that are sensitive to the profile shape: the concentration distribution with depth, surface concentration, integral mass gain, and signs of surface supersaturation. Additional process measurements, such as gas flow rate, pressure, and temperature logs, improve the observability of the hidden flux.

Regularization should include physical constraints and smoothness constraints. Nonphysical solutions are suppressed by penalties for negative flux, flux during the diffusion hold, exceeding the admissible surface concentration, and abrupt unstable changes in $J_C(t)$. In terms of inverse problems, this plays the role of prior information about the class of admissible boundary functions.

Model stability should be assessed not only by the error on the training database. It is necessary to test the sensitivity of the reconstructed flux to measurement noise, mesh changes, small temperature variations, and incomplete load data. An operator that sharply changes $J_C(t)$ in response to a small change in input parameters is unsuitable for technological use even if its error on the training regimes is low.

## 9. Calibration Functional and Training

The parameters θ are determined from the minimization problem:

$$\min_\theta L(\theta) = \lambda_C L_C + \lambda_m L_m + \lambda_s L_s + \lambda_p R_p + \lambda_u L_u .$$

The component $L_C$ describes the concentration-profile error, $L_m$ the error in the integral amount of introduced carbon, $L_s$ the surface-concentration error, $R_p$ physical regularization, and $L_u$ a term related to uncertainty or prediction stability.

$$L_C = \sum_i \left[ C_\theta\left(x_i, t_f\right) - C_{\exp}\left(x_i, t_f\right) \right]^2 ,$$

$$L_m = \left[ \Delta m_\theta - \Delta m_{\exp} \right]^2 ,$$

$$R_p = \int_0^{t_f} \left[ \alpha_1 \max\left(0, -J_C\right)^2 + \alpha_2 \left(1 - u\right) J_C^2 + \right.$$

$$\left. + \alpha_3 \max\left(0, C_s - C_{allow}\right)^2 + \alpha_4 \left( \frac{dJ_C}{dt} \right)^2 \right] dt .$$

Training can be performed in two variants. In the first variant, the numerical solver for the diffusion problem is retained, and the parameters θ are optimized through an external procedure analogous to training a model with a differentiable or nondifferentiable forward solver. In the second variant, a physics-informed approximation is used in which the residual of the diffusion equation directly enters the functional. For a materials-science implementation, the first variant is preferable at the initial stage because it preserves a verified numerical solver and localizes machine learning in the boundary operator.

## 10. Model-Construction Algorithm

The algorithm consists of two related procedures: preliminary calibration using a database of experimental regimes and online adaptation when the model is applied to a new process. The textual detailing of the steps has been replaced by flowcharts because, for the publication version, the graphical representation better shows the feedback structure.

### 10.1. Preliminary Calibration

A flowchart of the preliminary model-calibration algorithm is shown in Fig. 1.

**Algorithm 9.1. Preliminary Model Calibration**

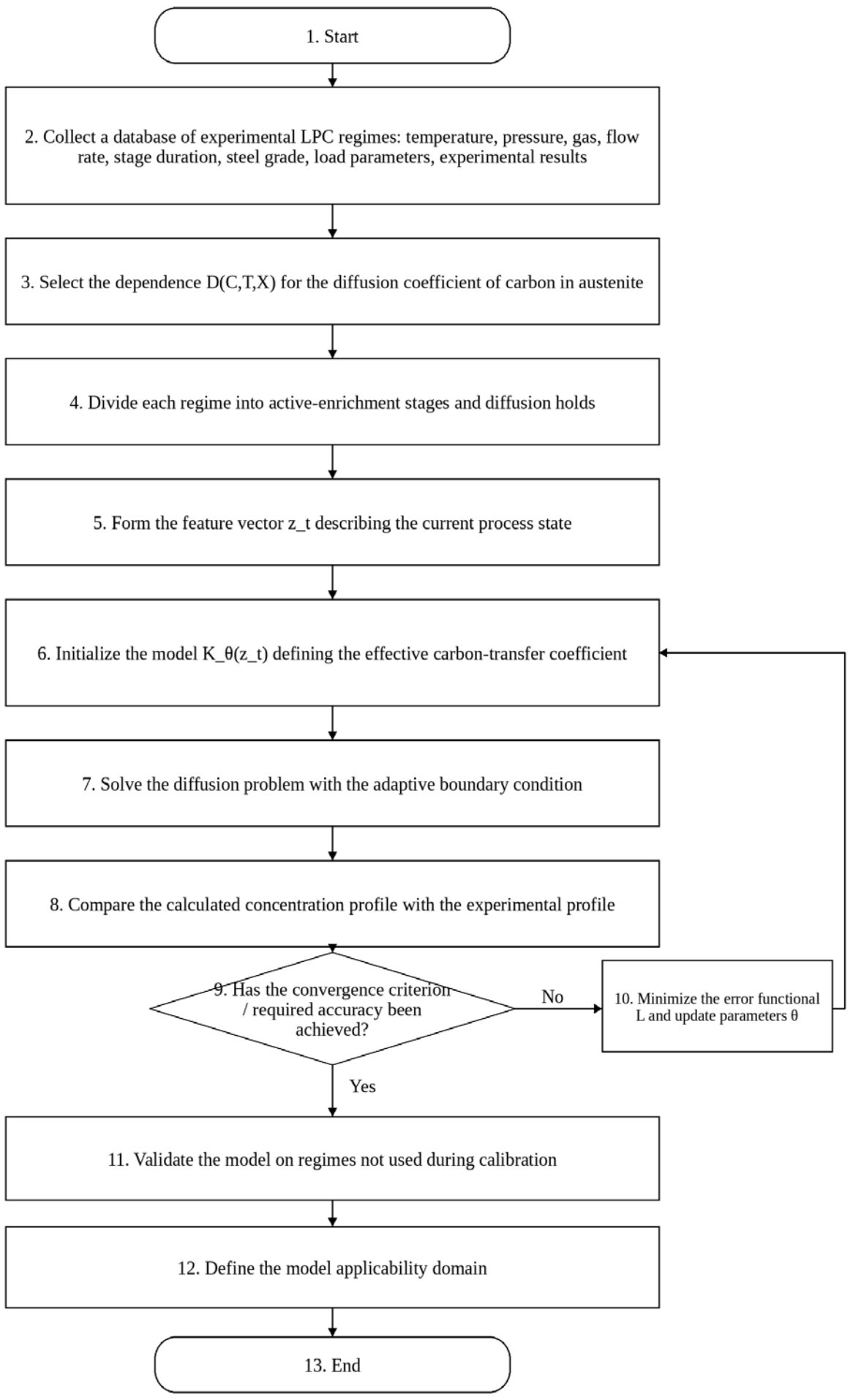

**Fig. 1.** Preliminary model-calibration algorithm

### 10.2. Online Adaptation

A flowchart of online model adaptation during execution of the technological process is shown in Fig. 2.

**Algorithm 9.2. Online Adaptation**

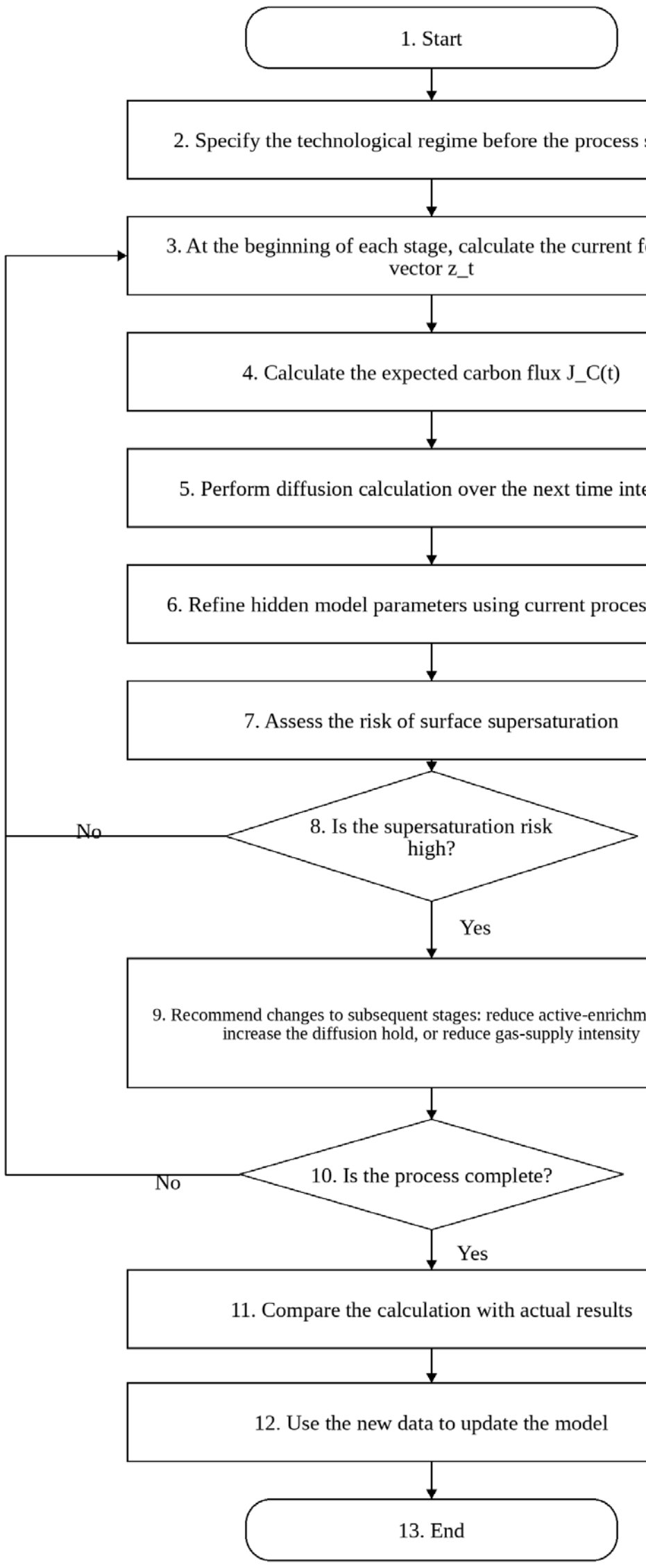

**Fig. 2.** Online model-adaptation algorithm

## 11. Model Verification and Quality Criteria

Validation should be performed by comparison with several baseline approaches: a model with a constant mean flux, a model with a predefined segmented flux, an empirical transfer coefficient, a direct "regime-result" model, and the proposed model with an adaptive boundary operator. This comparison is necessary to show that the gain is achieved specifically through the correct reconstruction of boundary kinetics and not merely through an increased number of fitting parameters.

The main quality criteria are the concentration-profile error, surface-carbon error, effective-case-depth error, integral mass-gain error, robustness to noise, transferability between regimes, and correctness of uncertainty estimation. The ability of the model to transfer to new sequences of active-enrichment stages and diffusion holds should be checked separately.

A minimal numerical example for the first verification should include a synthetic or experimental set of regimes with a known final profile. For each regime, it is necessary to compare the reconstructed $J_C(t)$, the calculated profile $C(x, t_f)$, and the integral mass balance. Such verification will make it possible to separate the identification of the boundary operator itself from errors in the diffusion part of the model.

## 12. Discussion

The proposed approach occupies an intermediate position between classical physical modeling and fully empirical machine learning. Its strength is that the transport physics does not dissolve into a black box. Machine learning is applied only where the physical model is genuinely incomplete: in the description of boundary kinetics.

From a theoretical point of view, the model can be interpreted as learning under constraints imposed by a partial differential equation and with a trainable boundary operator. This distinguishes it from standard PINNs, where the neural network often approximates the entire solution field. In the formulation considered here, the numerical solver and the physical model are retained, while the trainable object is the operator that links the technological trajectory with the boundary flux.

Practical applicability will be determined by data quality. If only final values of case depth are available, the reconstruction of $J_C(t)$ will be weakly identifiable. If concentration profiles, mass balance, load parameters, and process logs are available, the trainable operator receives enough information for physically meaningful calibration.

The proposed model does not eliminate the need for experimental verification. On the contrary, its main scientific meaning is the formulation of a testable hypothesis: adaptive reconstruction of the boundary flux should provide better transferability than a fixed or averaged flux while preserving the physical structure of the diffusion calculation.

## 13. Conclusions

An extended hybrid LPC model has been formulated in which carbon diffusion in austenite is defined by a physical transport equation, whereas the carbon flux through the surface is determined by a self-calibrating boundary operator.

It has been shown that the boundary flux is the natural location for applying machine learning: it is weakly observable, depends on many technological parameters, and directly affects the concentration profile, while also having to satisfy strict physical constraints.

The proposed formulation is related to inverse problems for transport equations, physics-informed learning, operator learning, and dynamic-system identification. Unlike direct empirical models, it preserves the physical diffusion operator and localizes uncertainty in the boundary closure.

The practical value of the approach should be verified by comparison with baseline flux models, analysis of transferability between regimes, and uncertainty quantification. If successfully validated, this formulation can serve as the basis for developing a computational tool for digital support of LPC; however, its domain of applicability must be defined explicitly and limited to the range of steels, temperatures, gas regimes, and load types represented in the calibration database.

### Conflict of interest

The authors declare that they have no conflict of interest in relation to this research, including financial, personal, authorship, or any other nature that could affect the research and its results presented in this article.

### Use of artificial intelligence

The authors confirm that they did not use artificial intelligence technologies when creating the current work.

## Фізико-інформаційне моделювання процесу цементації низького тиску (LPC): самокаліброванa гранична умова потоку вуглецю

**О. В. Сагалович[1] • Т. О. Сагалович[2] • С. Ф. Дуднік[1] • В. В. Сагалович[1]**

[1] *АТ "ФЕД" Харків, Україна*
[2] *Університет Прикладних наук і мистецтва, Люцерн, Швейцарія*

***Анотація.*** *У статті сформульовано розширену постановку гібридної фізико-інформаційної моделі цементації низького тиску (low-pressure carburizing, LPC), у якій дифузійне перенесення вуглецю в аустеніті описується фізичним рівнянням, а ефективний потік вуглецю через межу газ-метал розглядається як невідомий граничний вплив, що підлягає ідентифікації за експериментальними та технологічними даними. Підставою для такої постановки є те, що в LPC найбільш невизначеною частиною моделі є не об'ємна дифузія, а нестаціонарна поверхнева кінетика надходження вуглецю в метал. На відміну від прямих емпіричних залежностей типу "режим – результат", запропонована схема використовує машинне навчання як замикаючий оператор граничної умови в дифузійній задачі. Такий оператор калібрується за профілями концентрації вуглецю, інтегральним приростом маси, поверхневою концентрацією, ефективною глибиною шару та додатковими технологічними вимірюваннями. Теоретично модель позиціонується як задача навчання оператора в системі з обмеженнями, заданими рівнянням у частинних похідних, фізичними обмеженнями на потік, регуляризацією оберненої задачі та оцінюванням невизначеності. Розглянуто зв'язки запропонованого підходу з фізично інформованими нейронними мережами, операторним навчанням, ідентифікацією динамічних систем, навчанням граничних операторів і задачами керування на межі. Сформульовано математичну постановку, функціонал калібрування, умови ідентифікованості, алгоритм попереднього налаштування та схему поточної адаптації. Запропонований підхід розглядається як обчислювальна основа для розроблення інтерпретованого цифрового супроводу процесів LPC.*